\documentclass[aps,prl,twocolumn,showpacs]{revtex4}
\usepackage{graphicx}
\usepackage{amssymb}

\begin{document}

\title{Universality of the melting curves for a wide range of interaction potentials}

\author{Sergey A. Khrapak, Manis Chaudhuri, and Gregor E. Morfill}

\affiliation{Max-Planck-Institut f\"ur extraterrestrische Physik, D-85741 Garching,
Germany}

\date{\today}

\begin{abstract}
We demonstrate that the melting curves of various model systems of interacting particles collapse to (or are located very close to) a universal master curve on a plane of appropriately chosen scaled variables. The physics behind this universality is discussed. An equation for the emerging ``universal melting curve'' is proposed. The obtained results can be used to approximately predict melting of   various substances in a wide range of conditions.
\end{abstract}

\maketitle

Exploring phase behavior of different substances is an outstanding physical problem with significant impact on basic (e.g. theories of phase transitions) and applied (e.g. materials science) research.   Significant progress in this area has been achieved over the last decades using a variety of theoretical, experimental, and computational methods. This resulted in detailed phase diagrams of some conventional atomic and molecular substances, as well as different kinds of new materials such as fullerenes and soft matter (colloidal suspensions, polymers, surfactants, complex plasmas, etc.)~\cite{Likos,Malescio,ChaudhuriSM,FortovPR}.

The traditional procedure of {\it predicting} the phase diagram of a substance is based on extensive numerical simulations using a model pair potential which approximates actual interactions in this substance. Simple pair interactions studied in this context include the hard sphere, inverse power law, large family of Lennard-Jones-type, Yukawa, Gaussian and other model potentials~\cite{Hansen_Book,Hansen1970,Hoover1971,Agrawal,Hamaguchi,Prestipino2,LJ-type}.
There exists also several approximate methods to locate phase boundaries (especially for the fluid-solid phase changes), which are independent of an exact shape of the interaction. These are well known phenomenological criteria for freezing and melting, like e.g. Lindemann melting law, Hansen-Verlet freezing rule, Ravech\'{e}-Mountain-Street criterion for freezing, and a dynamical criterion for freezing in colloidal suspensions (for a review see Ref.~\cite{Lowen}). These criteria are typically based on the properties of only one of the two coexisting phases and predict quasi-universal values of certain structural or dynamical quantities. Quasi-universality means that a quantity is not exactly constant, but varies in a sufficiently narrow range for a broad variety of physical systems (i.e. interaction shapes). These useful empirical rules can sometimes be directly applied to the substance under investigation. More often, they are used in combination with numerical simulations, which allows to approximately locate phase boundaries with a very modest computational cost~\cite{Saija}. Nevertheless, it would be often desirable to have simple analytical expressions describing phase coexistence, based only on the properties of the interaction potential. Even though such expressions can be not completely universal (e.g. their applicability limited to a certain class of interactions), they can be quite helpful, especially in cases when interparticle interactions depend on a number of system parameters, which can vary from one situation to another~\cite{PRL2009}.

The main purpose of this Letter is to describe a universality of melting curves for a wide range of interparticle interactions. In particular, we identify a pair of scaled variables, which characterize the system of strongly interacting particles. We then show that in a plane of this variables the available numerical data for melting of quite different model systems are located on (or very close to) a single master curve. A simple expression for this ``universal melting curve'' is proposed. We briefly discuss this finding in the context of several relevant recent studies.

The physical idea is rather simple and transparent. Let us consider a system of particles interacting via a pairwise interaction potential exhibiting conventional properties: The interaction force is strongly repulsive at short separations and vanishes at infinite separations (the long-range asymptote can be either repulsive or attractive). We then assume that the particle system is {\it strongly coupled}, i.e. the system is either in a dense fluid or solid state. The particles form a regular structure where large deviations of the interparticle separation from its average value are very seldom. Consequently, the state of the system should be virtually insensitive to the exact shape of the interaction potential at short distances. If, in addition, the potential decays sufficiently fast for distances beyond the mean interparticle separation, the properties of the system are hardly influenced by the shape of the interaction potential at long distances. The behavior of the potential at the average interparticle separation plays a dominant role. Physically, the force and its derivative evaluated at the mean interparticle distance are the two quantities which should really matter. We therefore put forward the following hypothesis: Two systems of strongly interacting particles, characterized by different interaction potentials $U_1(r)$ and $U_2(r)$ should exhibit similar properties provided $U_1'(\Delta)=U_2'(\Delta)$ and $U_1''(\Delta)=U_2''(\Delta)$, where $\Delta$ is the mean interparticle distance. The purpose of this Letter is not to give a general proof of this hypothesis, but demonstrate that it leads to remarkably good results when applied to fluid-solid coexistence of various strongly coupled systems.

To proceed further it is useful to chose a reference potential of interaction. Since fluid-solid phase changes are dominated by the repulsive portions of the interparticle forces, the natural and simplest choice is the inverse-power-law (IPL) family of potentials $U(r)=\varepsilon (\sigma/r)^n$, where $\varepsilon$ and $\sigma$ are energy and length scales. This potential defines a system of ``soft'' spheres and is used to model e.g. one-component plasma (OCP) and simple metals under extreme thermodynamic conditions. The softness of the potential is governed by the index $n$ and can be smoothly varied in essentially entire range: the very soft $n=1$ repulsion corresponds to the OCP limit, the potential becomes harder as $n$ increases, and for $n\rightarrow \infty$ it approaches the hard sphere (HS) limit.

In the absence of external fields and dissipation the system of $N$ particles interacting via the IPL potential is described by Newton's equations of motion, which in reduced units have the form
\begin{equation}
\ddot{\bf r}_i={\mathcal F}\sum_{j\neq i}\frac{{\bf r}_i-{\bf r}_j}{|{\bf r}_i-{\bf r}_j|^{n+2}},
\end{equation}
where ${\mathcal F}=n(\varepsilon/T)(\sigma/\Delta)^{n}\equiv nT_*^{-1}\rho_*^{n/3}$ characterizes the force of interaction at the mean interparticle distance, $\Delta=(V/N)^{1/3}$ is the structure-independent interparticle spacing, $V$ is the system volume, $T_*=T/\varepsilon$ is the reduced temperature and $\rho_*=N\sigma^3/V$ is the reduced density. The single scaled variable ${\mathcal F}$ [or any other combination of $\rho_*^{n/3}/T_*$ with/without $n$] describes the whole range of thermodynamic properties of the given IPL potential.

\begin{figure}
\includegraphics[width=6.6cm]{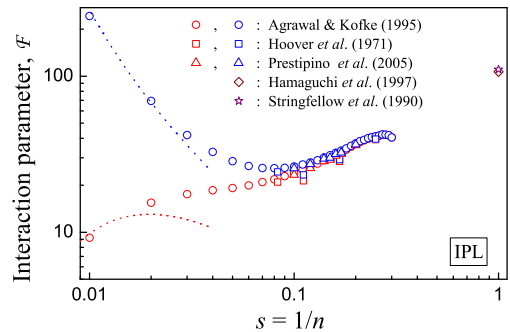}
\caption{(Color) Phase diagram of the IPL system in the $\mathcal F$ vs $1/n$ plane, where ${\mathcal F}=n(\varepsilon/T)(\sigma/\Delta)^{n}$ is the interaction parameter. Here red (blue) squares, circles, and triangles correspond to the representative numerical data for freezing (melting) taken from Refs.~\cite{Hoover1971},~\cite{AgrawalMolPhys} and~\cite{Prestipino}, respectively. Diamond and star mark the OCP ($n=1$) fluid-solid phase change as estimated in Ref.~\cite{Hamaguchi} and \cite{Stringfellow}, respectively. Red (blue) dashed curves correspond to the hard-sphere asymptotes ($n\rightarrow\infty$) for freezing (melting), see text.} \label{f1}
\end{figure}

The phase diagram of the IPL system has been extensively studied~\cite{Hoover1971,Agrawal,AgrawalMolPhys,Prestipino}. Three phases can exist, one fluid and two solid forming either a body-centered cubic (bcc) lattice, favorable for soft interactions ($n\lesssim 6$), or a face-centered cubic (fcc) lattice, favorable for hard interactions ($n\gtrsim 6$) \cite{Malescio,Agrawal}. Since we are mostly interested in the fluid-solid transition here, the phase diagram of the IPL system in the ($1/n$, ${\mathcal F}$) plane plotted in Fig.~\ref{f1} shows only the numerical data for freezing and melting.

The data in Fig.~\ref{f1} are from numerical simulations of Refs. \cite{Hoover1971,AgrawalMolPhys,Prestipino} and cover the wide range of softness between $n\simeq 3$ and $n=100$. Note that when $n$ approach 3 from above (in three-dimensional case under consideration), the difference between freezing and melting densities vanishes and melting and freezing curves fall on top of each other. When $n\leq 3$ the thermodynamic functions of the IPL system are undefined, unless a uniform ``neutralizing background'' is provided. In this case the fluid-solid phase change occurs at constant density (not at constant pressure), and freezing and melting points would coincide in the plane ($1/n$, ${\mathcal F}$). Two relevant simulation points, corresponding to the OCP regime at $n=1$ are also shown in Fig.~\ref{f1}.

Having defined and discussed the properties of the reference potential, let us identify the convincing pair of scaled variables, which will be used in further analysis.
We require that the first and the second derivative of some arbitrary potential $U(r)$ and the IPL potential are equal at $r=\Delta$. This immediately yields $n=-1-U''(\Delta)\Delta/U'(\Delta)$, which is one of the possible definitions of a local effective IPL exponent~\cite{Bailey}. For a given $n$ the phase state of the IPL system is fully determined by the value of the parameter ${\mathcal F}=-U'(\Delta)\Delta/T$. This provides us with the required pair of scaled variables: the {\it generalized softness parameter}
\begin{equation}\label{softness}
s=\left[-1-\frac{U''(\Delta)\Delta}{U'(\Delta)}\right]^{-1}
\end{equation}
and the {\it generalized interaction (force) parameter}
\begin{equation}\label{force}
{\mathcal F}=-U'(\Delta)\Delta/T.
\end{equation}
Our hypothesis can now be slightly reformulated in the form of the approximate corresponding states principle: Two different systems of strongly interacting particles having the same values of $s$ and ${\mathcal F} $behave alike. An ideal way to verify the vitality of this principle is to study the fluid-solid coexistence of various physical systems.

\begin{figure}
\includegraphics[width=6.6cm]{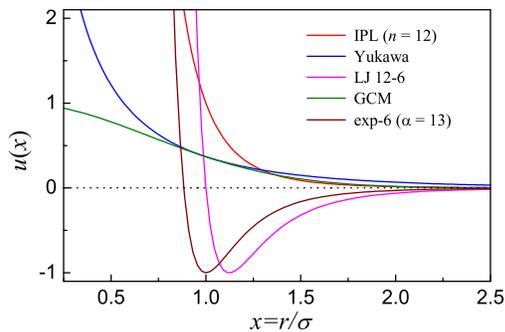}
\caption{(Color) Representative set of pair interaction potentials considered in this Letter. See text for description.
} \label{f2}
\end{figure}

We consider a wide range of interactions including purely repulsive and those having long-range attractive branches. It is convenient to present the potentials in the general form $U(r)=\varepsilon u(x)$, where $x=r/\sigma$ is the normalized distance. In this notation we have $u(x) = x^{-n}$ for the IPL potentials. Another purely repulsive potential is the Yukawa (Debye-H\"{u}ckel) potential $u(x)= (1/x)\exp(-x)$ extensively used in the context of colloidal suspensions, conventional electron-ion plasmas, and complex (dusty) plasmas~\cite{Likos,Malescio,ChaudhuriSM,FortovPR}. One of the most widely studied model potential capturing much of the essential physics of simple atomic substances is the $12-6$ Lennard-Jones (LJ) potential, $u(x) = 4(x^{-12}-x^{-6})$. There are also several modifications to the conventional $12-6$ LJ potential, including $n-6$ family of potentials,
\begin{displaymath}
u(x)=[n/(n-6)](n/6)^{6/(n-6)}(x^{-n}-x^{-6}),
\end{displaymath}
where $6<n<12$. The $\exp -6$ potential is defined as
\begin{displaymath}
u(x) = \left\{
\begin{array}{ll}
+ \infty, & ~ x < x_{\rm m} , \\ \frac{6}{\alpha -6}\exp\left[\alpha\left(1-x\right)\right]-\frac{\alpha}{\alpha -6} x^{-6}, & ~x\geq x_{\rm m},
\end{array} \right.
\end{displaymath}
where $x_{\rm m}$ corresponds to the maximum of the function in the second line, the repulsion is of ``hard sphere'' type for $x<x_{\rm m}$ and is controlled by the coefficient $\alpha$ for $x\geq x_{\rm m}$. The Weeks-Chandler-Anderson (WCA) reference repulsive potential correctly describes the repulsive forces in the $12-6$ LJ fluid for $x\leq 2^{1/6}$ and is set equal to zero at longer distances ~\cite{WCA}. One more potential we consider has a Gaussian form, $u(x)=\exp(-x^2)$ and is referred to as the Gaussian core model (GCM)~\cite{Stillinger}. It is widely used to describe effective interactions in soft matter physics (e.g polymers in solutions)~\cite{Likos}. Note that the bounded GCM potential does not diverge at $x\rightarrow 0$ as do the other potentials considered here. We have plotted a representative set of pair interaction potentials in Fig.~\ref{f2}.

\begin{figure}
\includegraphics[width=6.6cm]{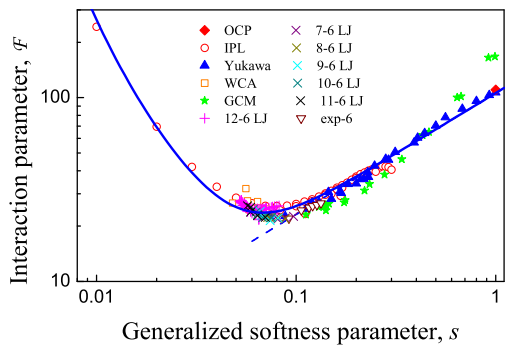}
\caption{(Color) The ``universal melting curve'' in the ($s$, ${\mathcal F}$) plane. Symbols (defined in the figure) denote the available numerical data. The solid curve is the fit of Eq.~(\ref{melting_fit}). } \label{f3}
\end{figure}

Figure \ref{f3} summarizes numerical data related to melting of the model potentials discussed above in the ($s$, ${\mathcal F}$) plane. The data for the IPL potential are the same as in Fig.~\ref{f1}, data for the Yukawa potential are from Refs.~\cite{Hamaguchi,Meijer,Stevens}, for the $12-6$ LJ potential we use numerical data from Refs.~\cite{Hansen1970,HanVerl_69,AK_95,BF_02,AS_09}, for the $n-6$ potential we use data from \cite{AS_09}, $\exp-6$ data for $\alpha=13$ are from \cite{SaijaPRB},  WCA data are from \cite{Kuijper}, and the data for the GCM potential are from Refs.~\cite{Prestipino,Mausbach}. Amazingly, most of the data points fall on (or very close to) a single curve -- the ``universal melting curve''. No systematic deviations are evident, except for the case of GCM potential (although even in this case data points are still located in the proximity of the universal curve). This is not surprising since GCM has a soft repulsive core and, therefore, for high densities one may expect some dependence of the system properties on the exact shape of the potential at short distances (absent for diverging repulsion).

We can propose a simple fit for the emerging ``universal melting curve''. For sufficiently soft interactions [soft spheres (SS) with $0.1\lesssim s\lesssim 1$] the data points can be reasonably described by a simple power-law dependence ${\mathcal F}_{\rm SS}\simeq 106 s^{2/3}$ (dashed line in Fig.~\ref{f3}), where the front factor 106 is taken to agree with the OCP ($s=1$) value from Ref.~\cite{Hamaguchi}. In the opposite limit of HS-like interaction ($s\rightarrow 0$) we use the following arguments (applied previously to GCM, e.g.~\cite{Stillinger}). The effective hard-sphere diameter $x_*$ can pe approximated as a distance at which the pair-interaction Boltzmann factor $\exp[-\varepsilon u(x)/T]$ is equal to $1/2$, i.e. $u(x_*)=T_*\ln 2$. The HS volume fraction at melting is $(\pi/6)x_*^3\rho_*\simeq 0.545$ ($\simeq 0.494$ at freezing). For the IPL system ${\mathcal F}=n \rho_*^{n/3}/T_*$, which immediately yields ${\mathcal F}\simeq n (1.041)^{n/3}\ln 2$ at melting [similarly ${\mathcal F}\simeq n (0.943)^{n/3}\ln 2$ at freezing]. This asymptotic estimates for melting and freezing are shown in Fig.~\ref{f1} by the dotted curves. They approach numerical data at low softness ($s\lesssim 0.02$). Generalizing the IPL case to other interactions we get the following HS asymptote for the melting curve ${\mathcal F}_{\rm HS}\simeq (\ln2/s)(1.041)^{1/3s}$. The simplest form which is correct in the corresponding limiting cases and yields reasonable interpolation between them is
\begin{equation}\label{melting_fit}
{\mathcal F} = \left({\mathcal F}_{\rm SS}^{\nu}+{\mathcal F}_{\rm HS}^{\nu}\right)^{1/\nu}.
\end{equation}
Equation (\ref{melting_fit}) with $\nu = 8/5$ do a remarkably good job (see Fig.~\ref{f3}).

Let us briefly discuss the observed universality in the context of some previous studies related to this topic. It was reported in Ref.~\cite{PRL2009} that the freezing curves of Yukawa and IPL potentials exhibit striking similarity when plotted in the plane of appropriately chosen parameters (the same would apply for the melting curves, especially when the interactions are sufficiently soft, so that freezing and melting curves are almost indistinguishable). The pair of generic parameters identified in Ref.~\cite{PRL2009} is the actual coupling strength [$\Gamma= U(\Delta)/T$] and the steepness of the interaction potential at the mean interparticle distance [$\gamma= |d\ln U(r)/d\ln r|_{r=\Delta}$]. This choice comes from requiring the potential itself and its first derivative to be equal to those of the IPL potential at $r=\Delta$, although this was not stated explicitly in \cite{PRL2009}. Our present approach based on the equality of the force and its first derivative is more physically sound (for instance, it allows us to deal with attractive interactions) and is thus superior to that of \cite{PRL2009}. However, it is easy to see that for Yukawa and IPL potentials both approaches yield similar results. Let us require that $\Gamma$ and $\gamma$ are equal for Yukawa and IPL potentials. Then ${\mathcal F}$ are equal too, but the generalized softness parameters $s$ are different by the factor $n^2/(n^2-n+1)$, which is 1 for $n=1$, tends to 1 for large $n$, and reaches the maximum value of $\simeq 1.3$ at $n=2$. This maximum discrepancy between the two approaches corresponds to the regime where no numerical data for the IPL potential are presently available ($3>n>1$).

Accurate approximations for the dependence of temperature on density along the freezing and melting curves of the 12-6 LJ fluid have been recently proposed~\cite{LJ_acc}. Melting equation has the form $T_*\simeq 1.97\rho_*^4-1.08\rho_*^2$. It is essentially ``exact'' in the high temperature limit and at the triple point and shows remarkably good agreement with the numerical data in the intermediate region. Similar equation can be easily derived on the basis of the present consideration. Note that the softness of the 12-6 LJ potential varies in a relatively narrow range between $\frac{1}{12}\simeq 0.083$ in the high-temperature limit and $\simeq 0.053$ at the triple point. In this range the universal melting curve exhibits minimum and the dependence of ${\mathcal F}$ on $s$ is rather weak. In the first approximation we can simply assume ${\mathcal F}\simeq {\rm const}$, where ${\rm const}\simeq 24$ (see Fig.~\ref{f3}). For the 12-6 LJ potential this yields $T_*\simeq 2.0\rho_*^4-1.0\rho_*^2$, in reasonable agreement with a more involved consideration~\cite{LJ_acc}.


Finally, we would like to point out that the reported universality of melting curves can be considered in the context of the so-called ``isomorph'' concept (see Ref.~\cite{Gnan} and references therein). In particular, we have seen that melting of a wide class of strongly coupled systems can be well reproduced by a properly chosen IPL reference system. More generally, our results are not inconsistent with the prediction that there exists a class of fluids which exhibit equivalence principle (in the sense that the system of isomorphic curves of one fluid can be mapped onto that of the other fluid) and that the fluid-solid coexistence curve is itself an isomorph~\cite{Gnan}.

To summarize, we have shown that the melting curves for a wide range of different potentials exhibit a universal shape on a plane of appropriately chosen reduced variables. These are the normalized force of interparticle interaction and the generalized softness parameter, both evaluated at the mean interparticle distance. Available numerical data for the melting of IPL, Yukawa, WCA, 12-6 LJ, $n-6$ LJ, $\exp -6$ systems are essentially collapsing on a single curve, data for the bounded GCM potential are not exactly on the curve, but in its proximity. We propose an analytical expression for this ``universal melting curve''.  Although this phenomenological melting equation cannot replace the proper thermodynamic prescription of the phase portraits of different systems, we believe that it can be very useful in approximately predicting, with very little effort, the melting transitions for various substances over a wide range of conditions.

This work was partly supported by DLR under Grant 50WP0203.


\begin{thebibliography}{99}


\bibitem{Likos} C. N. Likos, Phys. Reports {\bf 348}, 267 (2001).
\bibitem{Malescio} G. Malescio, J. Phys.:Condens. Matter {\bf 19}, 073101 (2007).
\bibitem{ChaudhuriSM} M. Chaudhuri {\it et al}., Soft Matter {\bf 7}, 1287 (2011).
\bibitem{FortovPR} V. E. Fortov {\it et al}.,  Phys. Reports {\bf 421}, 1 (2005).

\bibitem{Hansen_Book} J.-P. Hansen and I. R. McDonald, {\it Theory of simple liquids} (Elsevier, USA, 2006).
\bibitem{Hansen1970} J.-P. Hansen, Phys. Rev. A {\bf 2}, 221 (1970).
\bibitem{Hoover1971} W. G. Hoover, S. G. Gray, and K. W. Johnson, J. Chem. Phys. {\bf 55}, 1128 (1971).
\bibitem{Agrawal} R. Agrawal and D. A. Kofke, Phys. Rev. Lett. {\bf 74}, 122 (1995).
\bibitem{Hamaguchi} S. Hamaguchi, R. T. Farouki, and D. H. E. Dubin, Phys. Rev. E {\bf 56}, 4671 (1997).
\bibitem{Prestipino2} S. Prestipino, F. Saija, and P. V. Giaquinta, Phys. Rev. E {\bf 71}, 050102(R) (2005).
\bibitem{LJ-type} S. A. Khrapak {\it et al}.,
J. Chem. Phys. {\bf 134}, 054120 (2011).

\bibitem{Lowen} H. L\"{o}wen, Phys. Rep. {\bf 237}, 249 (1994).

\bibitem{Saija} F. Saija, S. Prestipino, and P. V. Giaquinta, J. Chem. Phys. {\bf 124}, 244504 (2006).

\bibitem{PRL2009} S. A. Khrapak and G. E. Morfill, Phys. Rev. Lett. {\bf 103}, 255003 (2009).

\bibitem{AgrawalMolPhys} R. Agrawal and D. A. Kofke, Mol. Phys. {\bf 85}, 23 (1995).
\bibitem{Prestipino} S. Prestipino, F. Saija, and P. V. Giaquinta, J. Chem Phys. {\bf 123}, 144110 (2005).

\bibitem{Stringfellow} G. S. Stringfellow, H. E. DeWitt, and W. L. Slattery, Phys. Rev. A {\bf 41}, 1105 (1990).

\bibitem{Bailey} N. P. Bailey {\it et al}., J. Chem. Phys. {\bf 129}, 184508 (2008).





\bibitem{WCA} J. D. Weeks, D. Chandler, and H. C. Andersen, J. Chem. Phys. {\bf 54}, 5237 (1971).
\bibitem{Stillinger} F. H. Stillinger, J. Chem. Phys. {\bf 65}, 3968 (1976).

\bibitem{Meijer} E. J. Meijer and D. Frenkel, J. Chem. Phys. {\bf94}, 2269 (1991).
\bibitem{Stevens} M. J. Stevens and M. O. Robbins, J. Chem. Phys. {\bf 98}, 2319 (1993).

\bibitem{HanVerl_69} J. P. Hansen and L. Verlet, Phys. Rev. {\bf 184}, 151 (1969).
\bibitem{AK_95} R. Agrawal and D. A. Kofke, Mol. Phys. {\bf 85}, 43 (1995).
\bibitem{BF_02} M. A. Barroso and A. L. Ferreira, J. Chem. Phys. {\bf 116}, 7145 (2002).
\bibitem{AS_09} A. Ahmed and R. J. Sadus, J. Chem. Phys. {\bf 131}, 174504 (2009).

\bibitem{SaijaPRB} F. Saija and S. Prestipino, Phys. Rev. B {\bf 72}, 024113 (2005).

\bibitem{Kuijper} A. de Kuijper, J. A. Schouten, and J. P. J. Michels, J. Chem. Phys. {\bf 93}, 3515 (1990).

\bibitem{Mausbach} P. Mausbach, A. Ahmed, R. Sadus, J. Chem. Phys. {\bf 131}, 184507 (2009).


\bibitem{LJ_acc} S. A. Khrapak and G. E. Morfill, J. Chem. Phys. {\bf 134}, 094108 (2011).

\bibitem{Gnan} N. Gnan {\it et al}., J. Chem. Phys. {\bf 131}, 234504 (2009).








%
%


%
%




%


\end{thebibliography}
\end{document}